\documentclass{article}
\usepackage{spconf,amsmath,graphicx}
\usepackage{cite}
\usepackage{overpic}
\usepackage{amsmath,amssymb,amsfonts}
\usepackage{graphicx}
\usepackage{textcomp}
\usepackage{xcolor}
\usepackage{float}
\usepackage{amsthm}
\usepackage{graphicx}
\usepackage{epstopdf}
\usepackage{amsmath,bm}
\usepackage{amsfonts}
\usepackage{amssymb}
\usepackage{color}
\usepackage{multirow}
\usepackage{multicol}
\usepackage{soul,xcolor}
\usepackage{algorithm}
\usepackage{algpseudocode}%

\theoremstyle{plain}

\newtheorem{lemma}{Lemma}

\newcommand{\vect}[1]{\mathbf{#1}}

\def\Htran{\mbox{\tiny $\mathrm{H}$}}
\def\Ttran{\mbox{\tiny $\mathrm{T}$}}
\def\imagunit{\mathsf{j}} 
\def\m{\rm}

\begin{document}

\title{A New Polar-Domain Dictionary Design for \\ the Near-field Region of Extremely Large Aperture Arrays}\vspace{-2cm}

\name{\"Ozlem Tu\u{g}fe Demir$^*$ and Emil Bj{\"o}rnson$^{\dagger}$\thanks{The work by Emil Bj\"ornson was supported by the FFL18-0277 grant from SSF.}}
\address{$^*$Department of Electrical-Electronics Engineering, TOBB ET\"U, Ankara, T\"urkiye \\
$^\dagger$Department of Computer Science, KTH Royal Institute of Technology, Kista, Sweden}

\ninept
\maketitle

\begin{abstract}
A grid of orthogonal beams with zero column coherence can be easily constructed to cover all prospective user equipments (UEs) in the far-field region of a multiple-antenna base station (BS). However, when the BS is equipped with an extremely large aperture array, the Fraunhofer distance is huge, causing the UEs to be located in the radiative near-field region. This calls for designing a grid of beams based on a near-field dictionary. In the previous work, a polar-domain grid design was proposed to maintain control over the column coherence. A limitation of this approach is identified in this paper, and we propose an enhanced methodology for the design of a polar-domain dictionary specifically tailored for the near-field of an extremely large aperture uniform planar array. Through simulation results, it is demonstrated that the proposed dictionary, employing a non-uniform distance sampling approach, achieves lower column coherence than the benchmark and significantly improves the localization of UEs compared to uniform distance sampling.
\end{abstract}


\section{Introduction}

Recently, there has been a notable shift towards extremely large aperture arrays, characterized by a significantly higher number of antennas compared to conventional massive MIMO (multiple-input multiple-output) base stations (BSs) and larger apertures. This paradigm shift aims to enhance spatial multiplexing, interference management, and beamforming gains in wireless communication systems \cite{hu2018beyond, Bjornson2019d, demir2022channel, ramezani2022near}.  In conventional arrays, the user equipments (UEs) are typically located in the far-field of the BS array. This allows the columns of a discrete Fourier transform matrix (DFT) to serve as an orthogonal basis, representing the array response vectors when employing uniform linear or planar arrays \cite{massivemimobook}. The far-field condition provides a dictionary of orthogonal beams with zero column coherence (i.e., the inner product magnitudes between columns), thereby simplifying beam training and management. However, in the context of extremely large aperture arrays, it is highly probable that some UEs will be located in the radiative near-field of the array, even in the sub-6\,GHz band \cite{bjornson2021primer,ramezani2022near}. Consequently, a novel analysis is required to construct a near-field codebook (dictionary) for efficient beam management in such scenarios \cite{you2023near}.

Unlike far-field beamforming, where the beamforming gain exhibits infinite depth along the range dimension for a given azimuth and elevation direction, the finite depth of near-field beamforming gain introduces the range domain as a novel design domain \cite{wu2023multiple,ramezani2022near,bjornson2021primer,cui2022near}. Recently, a polar-domain grid design has been proposed for the near-field of a uniform linear array (ULA) by keeping the column coherence under control, which is more challenging compared to its far-field counterpart \cite{cui2022channel}. Existing literature predominantly focuses on near-field dictionary design or beam management for ULAs \cite{zhang2022fast,zhang2023near,yang2023channel,wu2023parametric}. However, the utilization of uniform planar arrays (UPAs) offers distinct advantages in terms of accommodating many more antennas within a limited aperture area. \cite{wu2023multiple} extended the concept of polar-domain dictionary design to the near-field region of UPAs, considering the interplay of azimuth, elevation, and range dimensions. Due to the inherent complexity of the analysis, the authors addressed the angular and distance sampling independently to reduce column coherence along a single dimension. We have observed that this dictionary exhibits a high correlation among specific dictionary columns for a UPA operating in the sub-6 GHz band, despite the column coherence being small for most columns. This results in poor performance in localizing the grid point where the UE belongs. We postulate that the mismatch of the distance sampling among different angular directions is the cause of full column coherence among these columns.

In this paper, we present a novel methodology for polar-domain dictionary design in the near-field of an extremely large UPA. Our proposed approach employs a non-uniform and closed-form distance sampling method, which eliminates the need for function value search for each angular pair as required in \cite{wu2023multiple}, while simultaneously enabling control over the column coherence among different angular directions and distances. This effectively eliminates the issue of full correlation among distinct dictionary columns. 
Furthermore, we take into account the influence of arbitrary antenna spacing on the dictionary design, which is distinct from the existing approaches that assume a half-wavelength antenna spacing. The simulation results validate the efficacy of the proposed dictionary, utilizing the specialized non-uniform distance sampling approach, in achieving lower column coherence and significantly improving the localization of UE compared to the existing method and uniform sampling.

\section{System and Channel Modeling}

We consider the communication scenario involving a BS equipped with an extremely large aperture array and a UE with a single antenna. The $M$ BS antennas are deployed in the form of UPA. To illustrate the configuration, we refer to \cite[Fig.~1]{Bjornson2021b}, where the number of antennas per row and per column of the UPA is denoted as $M_{\rm H}$ and $M_{\rm V}$, respectively, resulting in $M=M_{\rm H}M_{\rm V}$. The spacing between adjacent antennas in the horizontal and vertical directions is $\Delta$. Our focus is on scenarios where the array consists of thousands of antennas and the spacing between antennas is less than half of the wavelength $\lambda$. The antennas are indexed in a row-by-row manner using the parameter $m$, ranging from $1$ to $M$. In accordance with the coordinate system in \cite[Fig.~1]{Bjornson2021b}, the position of the $m$th antenna relative to the origin is given by the vector $\vect{u}_m = [ 0, \, \,\, i(m) \Delta,  \,\,\, j(m) \Delta]^{\Ttran}$, where $i(m) =\mathrm{mod}(m-1,M_{\m H})$ and $j(m) =\left\lfloor(m-1)/M_{\m H}\right\rfloor$ are the horizontal and vertical indices of element $m$, respectively. Here, $\mathrm{mod}(\cdot,\cdot)$ is the modulus operation, while $\lfloor \cdot \rfloor$ denotes the truncation operation.

 The Fraunhofer (Rayleigh) distance $2D^2/\lambda$, where $D=\sqrt{M_{\rm H}^2+M_{\rm V}^2}\Delta$ is the aperture length, characterizes the classical boundary between the far-field and radiative near-field of an array. Since the UPA is extremely large, the Fraunhofer distance can be at order of hundreds of meters \cite{ramezani2022near}; hence, in this paper, we focus on the scenario where the UEs are in the (radiative) near-field region, Fresnel region, of the extremely large UPA and the channel between the BS and UE is line-of-sight (LOS). This means that the distance from the UPA to the UE antenna is less than Fraunhofer distance $2D^2/\lambda$ but larger than the Fresnel distance $0.62\sqrt{D^3/\lambda}$.

The LOS channel of an arbitrary UE in the near-field region of the array is denoted by $\vect{h}\in \mathbb{C}^M$. A wave coming from the near-field has a spherical wavefront, implying that the channel can be expressed as \cite{cui2022channel}
\begin{equation}
\vect{h}= \sqrt{\beta}\vect{b}\left(\varphi, \theta, r\right), 
\end{equation}
where $r$ is the distance between the origin (corner point of the UPA) and the UE antenna, and $\beta>0$ is the distance- and angle-dependent channel gain, which takes the path loss and antenna directivity into account. The near-field array response vector $\vect{b}\left(\varphi, \theta, r\right)$ also depends on the azimuth and elevation angles $\varphi$ and $\theta$, which are computed as seen from the origin. The near-field array response vector characterizes the specific spherical wave and is written for the considered UPA as
\begin{equation} \label{eq:near-field-array-response-vector}
\vect{b}\left(\varphi, \theta, r\right) = \left[e^{-\imagunit\frac{2\pi}{\lambda}(r_{1}-r)}, \ldots, e^{-\imagunit\frac{2\pi}{\lambda}(r_{M}-r)} \right]^{\Ttran}  \vspace{-2mm}
    \end{equation}
where $r_{m}$ denotes the distance from the BS antenna $m$ at the location $\vect{u}_m = [ 0, \, \,\, i(m) \Delta,  \,\,\, j(m) \Delta]^{\Ttran}$ to the UE antenna and it is given as 
 \vspace{-2mm}
\begin{align} \label{eq:distance-difference}
r_{m}=\bigg(&\Big(r\cos(\theta)\cos(\varphi)-0\Big)^2
+\Big(r\cos(\theta)\sin(\varphi)-i(m)\Delta\Big)^2\nonumber\\
&+\Big(r\sin(\theta)-j(m)\Delta\Big)^2\bigg)^{\frac12} \nonumber \\
&\hspace{-5.5mm}=r\Bigg( 1-2\Delta\frac{i(m)\cos(\theta)\sin(\varphi)+j(m)\sin(\theta)}{r} \nonumber\\
&\hspace{3mm}+\Delta^2\frac{i^2(m)+j^2(m)}{r^2}\Bigg)^{\frac12}.
\end{align}
We suppose that the waves arrive only from the directions in front of the array; that is, $\varphi \in [-\frac{\pi}{2},\frac{\pi}{2}]$. 

In the following section, we will propose a polar-domain dictionary design for LOS channels from the UPA with controllable column coherence. This dictionary will be useful for beam tracking and sparse channel estimation techniques.

\section{Polar-Domain Dictionary Design for UPA}

When the far-field is considered, the orthogonal beams constructed by the columns of the DFT matrix serve as a dictionary for beam tracking and compressed sensing-based channel estimation. In that case, the column coherence, which is defined as the maximum of the absolute inner products between two different columns of the dictionary matrix, is zero since the columns of the dictionary are mutually orthogonal. A small value of the column coherence is advantageous in several respects. It promotes sparsity, which enables accurate channel estimation in a sparse channel environment. In addition, it facilitates precise beam tracking.
In \cite{cui2022channel,wu2023multiple}, a polar-domain dictionary matrix $\vect{W}$ that has the near-field array response vectors as its columns is proposed to minimize the column coherence for a ULA and UPA, respectively. The term ``polar'' comes from the joint angular and distance sampling methodology when constructing the columns of $\vect{W}$, which correspond to the grid points in the on-grid sparse channel recovery or the beam directions in tracking applications. Following the compressed sensing framework \cite{bajwa2010compressed}, the columns of $\vect{W}$ are selected from potential near-field array response vectors $\vect{b}\left(\varphi, \theta, r\right)$ in \eqref{eq:near-field-array-response-vector} so that the column coherence is
\begin{equation} \label{eq:column-coherence}
    \mu=\underset{p\neq q}{\max} \left|\vect{b}^{\Htran}\left(\varphi_{p}, \theta_{p}, r_{p}\right) \vect{b}\left(\varphi_{q}, \theta_{q}, r_{q}\right)\right|  
\end{equation}
becomes as small as possible. In \eqref{eq:column-coherence}, $p$ and $q$ are the column indices of $\vect{W}$. To make the analysis tractable, we utilize $\sqrt{1+x}\approx 1+\frac{1}{2}x-\frac{1}{8}x^2$ for small $x$ to approximate $r_{m}$ in \eqref{eq:distance-difference} as \cite{cui2022channel,wu2023multiple}
 \vspace{-2mm}
\begin{align}  \label{eq:near-field-expansion}
&r_{m}\approx  r -\Delta\Big(i(m)\cos(\theta)\sin(\varphi)+j(m)\sin(\theta)\Big) \nonumber\\
&\hspace{-2mm}+\!\!\Delta^2\!\!\left(\!\frac{\!i^2(m)\!+\!j^2(m)\!-\!\Big(\!i(m)\cos(\theta)\sin(\varphi)\!+\!j(m)\sin(\theta)\!\Big)^2}{2r}\!\right)
\end{align}
where we have also omitted the terms with $1/r^2,\ldots, 1/r^4$. This approximation is called the \emph{near-field expansion} \cite{ziomek1993three} and includes more phase terms than the classical Fresnel approximation, which omits the last term in \eqref{eq:near-field-expansion}. When $r$ is beyond the Fraunhofer distance, the last two terms involving $1/r$ can be omitted, and the array response vector in \eqref{eq:near-field-array-response-vector} becomes identical to the corresponding far-field array response vector.
Let us shift our focus to the summation of the last two terms in \eqref{eq:near-field-expansion}, by writing it as
\begin{align}
&\Delta^2 \frac{i^2(m)\left(1-\cos^2(\theta)\sin^2(\varphi)\right)}{2r}+\Delta^2\frac{j^2(m)\left(1-\sin^2(\theta)\right)}{2r} \nonumber\\
&- \Delta^2\frac{i(m)j(m)\cos(\theta)\sin(\varphi)\sin(\theta)}{r}. \label{eq:three-terms}
\end{align}
To make the upcoming analysis tractable, we will omit the third term in \eqref{eq:three-terms} as done in \cite{wu2023multiple}, which showed that when $M$ is moderately large, this term can be neglected safely. As an alternative way, irrespective of the number of antennas, the following lemma provides a basis for this assumption by showing that the maximum of the two scalars multiplying the first and second terms is always greater than or equal to the scalar multiplying the last term, which will be omitted.

\begin{lemma} Denoting $\Phi=\cos(\theta)\sin(\varphi)$ and $\Omega=\sin(\theta)$, for any $\varphi, \theta \in [-\frac{\pi}{2},\frac{\pi}{2}]$, it holds that
   \vspace{-2mm}
\begin{equation}
  \max\left(1-\Phi^2,1-\Omega^2\right)\geq |\Phi\Omega|.  
\end{equation}

   \vspace{-4mm}
\begin{proof}
Without loss of generality, let us consider the case $|\Phi|\geq |\Omega|$. Noting that $\Phi^2+\Omega^2=1-\cos^2(\theta)\cos^2(\varphi)\leq 1$, it must hold that 
$\max\left(1-\Phi^2,1-\Omega^2\right)=1-\Omega^2\geq \Phi^2 \geq |\Phi\Omega|$,  which concludes the proof.
\end{proof}
\end{lemma}

Inserting the first two terms from \eqref{eq:three-terms} into \eqref{eq:near-field-expansion} and \eqref{eq:near-field-array-response-vector}, we obtain the proposed approximation of the $m$th entry of the near-field array response vector $\vect{b}(\varphi,\theta,r)$ as
\begin{align}  \label{eq:proposed-approximation}
&\exp\Bigg(\imagunit\frac{2\pi}{\lambda}\Bigg[\Delta  \Big(i(m)\cos(\theta)\sin(\varphi)+j(m)\sin(\theta)\Big) \nonumber\\
&\!-\!\Delta^2 \frac{i^2(m)\left(1\!-\!\cos^2(\theta)\sin^2(\varphi)\right) \!+ \!j^2(m)\left(1\!-\!\sin^2(\theta)\right)}{2r}\Bigg]\Bigg).
\end{align}

\begin{figure}[t!]
	\hspace{-1cm}
		\begin{center}
			\includegraphics[trim={0.1cm 0.1cm 1cm 0.3cm},clip,width=3in]{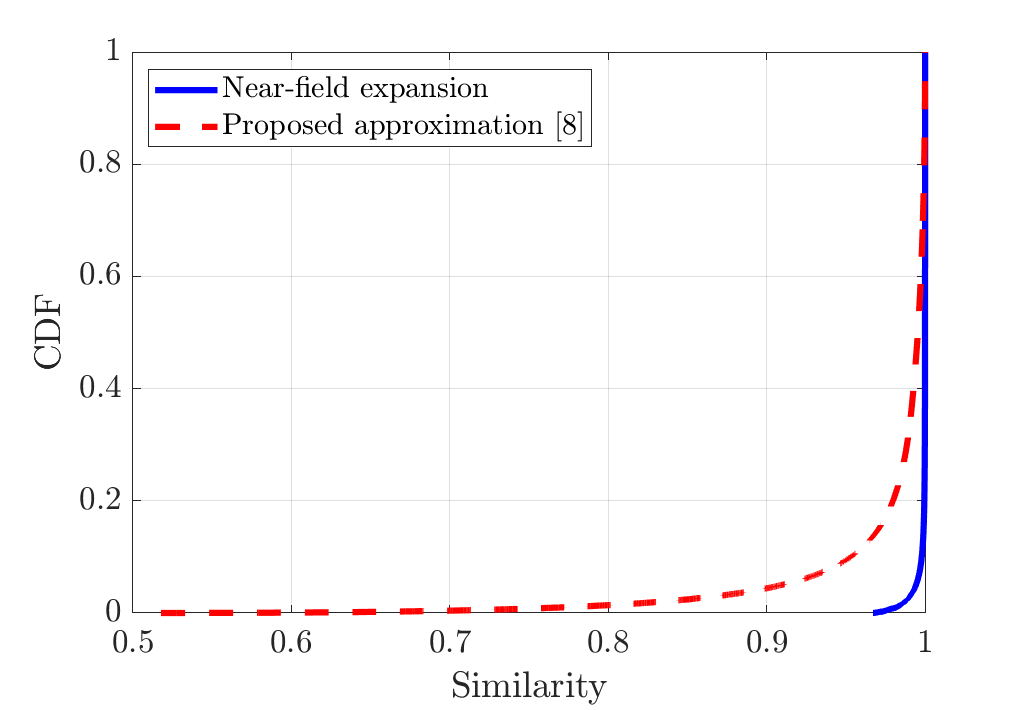}
   \vspace{-4mm}
			\caption{The CDF of the similarity (normalized absolute inner product) between the actual near-field array response vector in \eqref{eq:near-field-array-response-vector} and the two approximations: i) near-field expansion from \eqref{eq:near-field-expansion} and ii) the proposed approximation from \eqref{eq:proposed-approximation}.} \label{fig:similarity}
		\end{center}
     \vspace{-8mm}
\end{figure}

To quantify the accuracy of the considered approximation, we compute the similarity of the approximate near-field array response vector $\vect{b}_{\rm approx}(\varphi,\theta,r)$ to the actual $\vect{b}(\varphi,\theta,r)$ in \eqref{eq:near-field-array-response-vector} as
\vspace{-1mm}
\begin{equation}
    \mathrm{Similarity} =  \frac{\left|\vect{b}^{\Htran}_{\rm approx}(\varphi,\theta,r)\vect{b}(\varphi,\theta,r)\right|}{M} \label{eq:similarity},  \vspace{-1mm}
\end{equation}
which takes values between 0 and 1. In Fig.~\ref{fig:similarity}, we consider a  UPA with $M_{\rm H}=64$ antennas per row, $M_{\rm V}=32$ antennas per column, and the antenna spacing $\Delta=0.25\lambda$ where $\lambda=0.1$\,m (3\,GHz carrier frequency). The aperture length is $D=\sqrt{M_{\rm H}^2+M_{\rm V}^2}\Delta\approx1.79$\,m, which leads to the Fresnel distance of $0.62\sqrt{D^3/\lambda}\approx 4.69$\,m and the Fraunhofer distance of $2D^2/\lambda= 64$\,m. We uniformly sample $50$ azimuth angles and $50$ elevation angles in the range $[-\frac{0.9\cdot \pi}{2},\frac{0.9\cdot\pi}{2}]$, and 50 distance values in $[8,64]$\,m. We plot the cumulative distribution function (CDF) of the similarity values in \eqref{eq:similarity} computed for $50^3$ grid points. We consider two approximations: i) the near-field expansion from \eqref{eq:near-field-expansion} and ii) the proposed approximation from \eqref{eq:proposed-approximation}. The figure demonstrates that the near-field expansion provides almost full similarity for all the grid points whereas there are certain outliers that result in lower similarity values when considering the proposed approximation. However, the similarity is at least $0.9$ for more than $95\%$ of the locations. Based on that, we will continue with the proposed approximation in \eqref{eq:proposed-approximation} to design the polar-domain dictionary. In this case, the magnitude of the inner product in \eqref{eq:column-coherence} for a given pair of locations specified by $(\varphi_p,\theta_p,r_p)$ and $(\varphi_q,\theta_q,r_q)$ is computed in \eqref{eq:beamforming-gain} on the top of the next page, where we have defined $\Phi_q\triangleq \cos(\theta_q)\sin(\varphi_q)$, $\Phi_p\triangleq \cos(\theta_p)\sin(\varphi_p)$, $\Omega_q\triangleq \sin(\theta_q)$, and $\Omega_p\triangleq \sin(\theta_p)$. The summation is written as a multiplication of two separate summations over horizontal and vertical antenna dimensions as in \cite{wu2023multiple}. Thanks to the considered approximation, the terms $\Phi_q$, $\Phi_p$ only appear in the horizontal summation whereas the terms $\Omega_q$, $\Omega_p$ only appear in the vertical summation. On the other hand, the distances $r_q$, $r_p$ exist in both terms, which creates coupling and makes the design of polar-domain dictionary non-trivial.

\begin{figure*}
\begin{align} \label{eq:beamforming-gain}
   \left|\vect{b}^{\Htran}\left(\varphi_{p}, \theta_{p}, r_{p}\right) \vect{b}\left(\varphi_{q}, \theta_{q}, r_{q}\right)\right| \approx &\Bigg| \sum_{m=1}^{M_{\rm H}} \exp\Bigg(\imagunit\frac{2\pi}{\lambda}\Bigg[\Delta  (m-1)(\Phi_q-\Phi_p)-\Delta^2\frac{(m-1)^2(1-\Phi^2_q)}{2r_q} + \Delta^2\frac{(m-1)^2(1-\Phi^2_p)}{2r_p}\Bigg]\Bigg) \nonumber\\
&\times \sum_{n=1}^{M_{\rm V}} \exp\Bigg(\imagunit\frac{2\pi}{\lambda}\Bigg[\Delta  (n-1)(\Omega_q-\Omega_p)-\Delta^2\frac{(n-1)^2(1-\Omega^2_q)}{2r_q} + \Delta^2\frac{(n-1)^2(1-\Omega^2_p)}{2r_p}\Bigg]\Bigg)\Bigg|.
\end{align}
\hrulefill
   \vspace{-10mm}
\end{figure*}

\begin{figure}[t!]
	\hspace{-1cm}
		\begin{center}
			\includegraphics[trim={0.1cm 0.1cm 0.1cm 0.3cm},clip,width=3in]{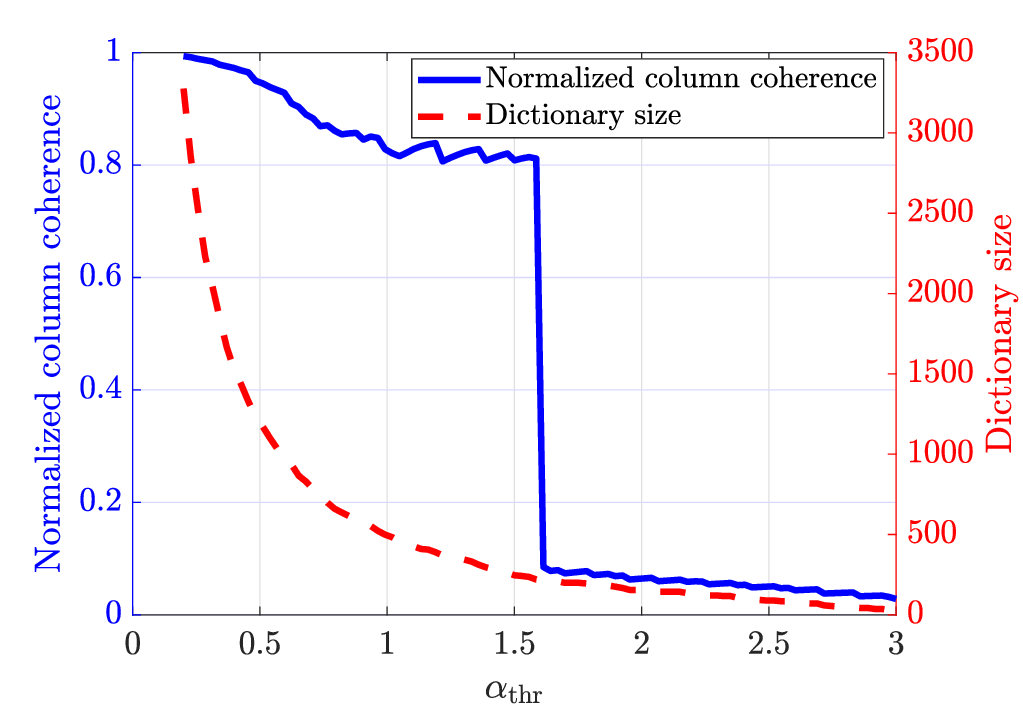}
      \vspace{-4mm}
			\caption{The normalized column coherence (left $y$-axis) and dictionary size (right $y$-axis) with respect to $\alpha_{\rm thr}$. } \label{fig:coherence}
		\end{center}
     \vspace{-8mm}
\end{figure}

\vspace{-3mm}

\subsection{Proposed Angular Sampling}
\vspace{-2mm}
For a given pair of azimuth and elevation angles, and, thus $(\Phi,\Omega)$, suppose the distances are sampled from the ``distance ring'' defined as
\begin{equation} \label{eq:distance-ring}
  \frac{(1-\Phi^2)(1-\Omega^2)}{r} = c ,
\end{equation}
where $c$ is a constant in generating the columns of the dictionary matrix $\vect{W}$. The above relation is the extension of the distance sampling for the case of ULAs  in \cite{cui2022channel} to UPAs and is different from the distance sampling proposed in \cite{wu2023multiple}. Consider two columns of the dictionary matrix with the distinct angular pairs $(\Phi_q,\Omega_q)$ and $(\Phi_p,\Omega_p)$ such that $\Phi_q\neq \Phi_p$ but $\Omega_q=\Omega_p$. If the respective distances in the dictionary are selected according to \eqref{eq:distance-ring}, then the magnitude of the inner product of these two columns from \eqref{eq:beamforming-gain} is proportional to
\vspace{-2mm}
\begin{align}
&\left| \sum_{m=1}^{M_{\rm H}} \exp\Bigg(\imagunit\frac{2\pi}{\lambda}\Bigg[\Delta  (m-1)(\Phi_q-\Phi_p)\Bigg]\Bigg)\right| \nonumber \\
=&  \left|\frac{\sin\left( M_{\rm H}\frac{\pi \Delta (\Phi_q-\Phi_p)}{\lambda} \right) }{ \sin \left(\frac{\pi \Delta (\Phi_q-\Phi_p)}{\lambda}\right)}\right|.
\end{align}
To make the above inner product zero, we should sample $\Phi=\cos(\theta)\sin(\varphi)$ so that
\vspace{-2mm}
\begin{align}\label{eq:phi-sampling}
    \Phi =  \frac{m\lambda }{M_{\rm H}\Delta}, \quad m=0,\pm 1, \pm 2, \ldots, \pm \left \lfloor \frac{M_{\rm H}\Delta}{\lambda}\right \rfloor.
\end{align}
Similarly, consider two columns of the dictionary matrix with distinct angular pairs $(\Phi_q,\Omega_q)$ and $(\Phi_p,\Omega_p)$ such that $\Phi_q=\Phi_p$ but $\Omega_q\neq \Omega_p$. If the respective distances in the dictionary are selected according to \eqref{eq:distance-ring}, then the magnitude of the inner product of these two columns from \eqref{eq:beamforming-gain} is proportional to
\begin{align}
& \left| \sum_{n=1}^{M_{\rm V}} \exp\Bigg(\imagunit\frac{2\pi}{\lambda}\Bigg[\Delta  (n-1)(\Omega_q-\Omega_p)\Bigg]\Bigg)\right| \nonumber \\
=& \left|\frac{\sin\left( M_{\rm V}\frac{\pi \Delta (\Omega_q-\Omega_p)}{\lambda} \right) }{ \sin \left(\frac{\pi \Delta (\Omega_q-\Omega_p)}{\lambda}\right)}\right|.
\end{align}
To make the above inner product zero, we should sample $\Omega=\sin(\theta)$ so that
\vspace{-2mm}
\begin{align} \label{eq:omega-sampling}
    \Omega =  \frac{n\lambda }{M_{\rm V}\Delta}, \quad n=0,\pm 1, \pm 2, \ldots, \pm \left \lfloor \frac{M_{\rm V}\Delta}{\lambda}\right \rfloor.
\end{align}
 We construct the angular grid from all possible $\Phi$ and $\Omega$ in \eqref{eq:phi-sampling} and \eqref{eq:omega-sampling}, which satisfy $\Phi^2+\Omega^2\leq 1$. This angular sampling matches with the far-field dictionary design and generalizes \cite{wu2023multiple} to support arbitrary antenna spacings.

   \vspace{-2mm}
\subsection{Proposed Distance Sampling}

To determine the distance sampling in line with \eqref{eq:distance-ring}, we 
    consider the two columns of the dictionary matrix with identical $\Phi_q=\Phi_p=\Phi$ and $\Omega_q=\Omega_p=\Omega$ values but with different distances $r_q\neq r_p$. In this case, we can approximate \eqref{eq:beamforming-gain} as \cite{wu2023multiple} 
    \begin{align} \label{eq:fresnel-integrals}
M\left|\frac{C(\alpha_{\rm H})+\imagunit S(\alpha_{\rm H})}{\alpha_{\rm H}}\right|  \left|\frac{C(\alpha_{\rm V})+\imagunit S(\alpha_{\rm V})}{\alpha_{\rm V}}\right|,
\end{align}
where $C(\alpha)=\int_{0}^{\alpha}\cos\left(\frac{\pi}{2}t^2\right)dt$ and $S(\alpha)=\int_{0}^{\alpha}\sin\left(\frac{\pi}{2}t^2\right)dt$ are the Fresnel integrals, $\alpha_{\rm H}=\sqrt{\frac{2M_{\rm H}^2\Delta^2(1-\Phi^2)}{\lambda}\left|\frac{1}{r_p}-\frac{1}{r_q}\right|}$, and $\alpha_{\rm V}=\sqrt{\frac{2M_{\rm V}^2\Delta^2(1-\Omega^2)}{\lambda}\left|\frac{1}{r_p}-\frac{1}{r_q}\right|}$.

As demonstrated in \cite[Fig.~6]{cui2022channel}, the function $\left|\frac{C(\alpha)+\imagunit S(\alpha)}{\alpha}\right|$ has an oscillating  pattern with decreasing values as $\alpha>0$ increases. Similarly, one can show that the approximate inner product magnitude 
 in \eqref{eq:fresnel-integrals} is likely to have smaller values when $\alpha=\alpha_{\rm H}\alpha_{\rm V}$ increases. Hence, it is possible to control the column coherence by setting a threshold $\alpha_{\rm thr}$ so that the sampled distances  for a given angular pair $(\Phi,\Omega)$  satisfy $\alpha=\alpha_{\rm H}\alpha_{\rm V}\geq\alpha_{\rm thr}$, i.e.,
 \begin{align} \label{eq:alpha-threshold}
     \alpha = \frac{2M_{\rm H}M_{\rm V}\Delta^2\sqrt{(1-\Phi^2)(1-\Omega^2)}}{\lambda}\left|\frac{1}{r_p}-\frac{1}{r_q}\right| \geq \alpha_{\rm thr}.
 \end{align}
It can easily be shown that if we sample the distances as
\begin{align} \label{eq:distance-sampling}
    r=  \frac{2M_{\rm H}M_{\rm V}\Delta^2}{\lambda\alpha_{\rm thr}}\frac{(1-\Phi^2)(1-\Omega^2)}{s}, \quad s=1,2,\ldots,
\end{align}
we both satisfy the previously defined distance sampling rule in \eqref{eq:distance-ring} and the condition in \eqref{eq:alpha-threshold}. This novel distance sampling is different from the one in  \cite{wu2023multiple}, which requires a search over values of a function for each angular pair $(\varphi,\theta)$ and does not guarantee \eqref{eq:distance-ring}.

To demonstrate the performance, we consider same simulation setup as in Fig.~\ref{fig:similarity} and construct a dictionary matrix with the angles sampled according to \eqref{eq:phi-sampling} and \eqref{eq:omega-sampling}, which satisfy $\Phi^2+\Omega^2\leq 1$ and the distances sampled according to \eqref{eq:distance-sampling} with varying $\alpha_{\rm  thr}$. We only keep the distances that are greater than or equal to $8$\,m. Fig.~\ref{fig:coherence} demonstrates the normalized column coherence obtained by dividing $\mu$ in \eqref{eq:column-coherence} by $M$, so that the value $1$ corresponds to full correlation. As $\alpha_{\rm thr}$ increases, except for slight oscillations, the normalized column coherence decreases. Around $\alpha_{\rm thr}=1.6$, a sharp decrease occurs. However, the bottleneck is the consistent reduction in the dictionary size, which is not desired. Hence, a balanced trade-off is required according to the needs of a particular application. 

   \vspace{-2mm}
\subsection{Comparison with Uniform Distance Sampling}
\vspace{-2mm}
 \begin{figure}[t!]
	\hspace{-1cm}
		\begin{center}
			\includegraphics[trim={0.1cm 0.1cm 1cm 0.3cm},clip,width=3in]{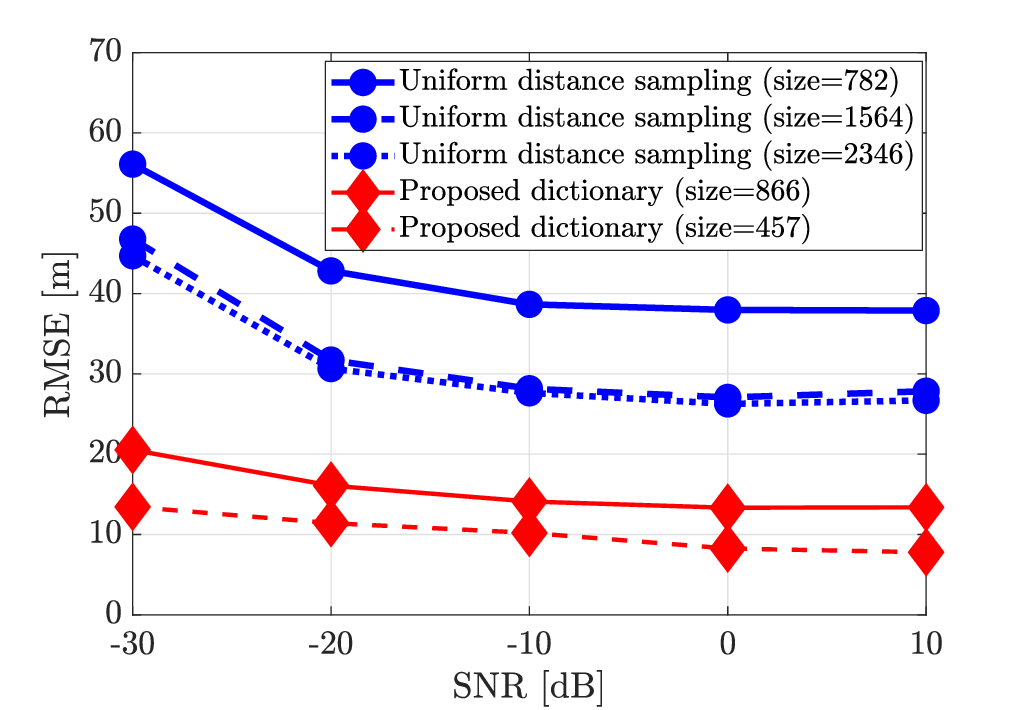}
      \vspace{-4mm}
			\caption{The RMSE between the actual and estimated grid locations for the proposed dictionary with the non-uniform distance sampling and uniform distance sampling.} \label{fig:performance}
		\end{center}
     \vspace{-8mm}
\end{figure}

An important practical application of the proposed dictionary, featuring reduced column coherence, is localizing the grid point closest to a UE's location. Of particular interest is the extent of performance enhancement achievable through the adoption of the proposed non-uniform distance sampling presented in \eqref{eq:distance-sampling}, as compared to the conventional uniform distance sampling method. To quantify the localization performance, we consider the previous simulation setup with the same angular sampling as in Fig.~\ref{fig:coherence}. Regarding distance sampling, we consider the proposed one with $\alpha_{\rm thr}=0.6525$ and $\alpha_{\rm thr}=1.0485$ selected from Fig.~\ref{fig:coherence} and three different uniform distance sampling with $2$, $4$, and $6$ grid points in $[8,64]$\,m. We randomly drop a UE with the azimuth and elevation angles in $[-\frac{0.9\cdot \pi}{2},\frac{0.9\cdot\pi}{2}]$, and distance in $[8,64]$\,m. We compute the nearest grid location to the UE from the dictionary and estimate this location by correlating the received noisy signal with each column of the dictionary matrix and selecting the maximum value. The root-mean-squared error (RMSE) between the actual and estimated grid locations in meters is obtained by averaging over $1000$ random UE locations for the different distance sampling methods. The RMSE is plotted in Fig.~\ref{fig:performance} with respect to the signal-to-noise ratio (SNR), and the dictionary size for each method is noted in the legend.  The figure shows that the proposed dictionary design with non-uniform distance sampling performs significantly better than uniform sampling. This is due to the smaller column coherence, which enables better discrimination of grid points. Through the simulations, we observed a much worse RMSE when considering the distance sampling rule in \cite{wu2023multiple} since although small, a certain number of dictionary columns have full correlation among them. Therefore, the achievement of fully discriminated dictionary columns through the proposed method ensures not only an acceptable but also significantly improved localization performance.

\vspace{-5mm}

\section{Conclusions}
\vspace{-3mm}

We have introduced a novel polar-domain dictionary design method for the near-field of a UPA. Our proposed dictionary is derived considering an arbitrary antenna spacing. We achieve lower column coherence by eliminating the mismatch between the distance and the angular sampling criterion. The proposed dictionary results in strictly lower column coherence than full coherence, thus fully discriminating dictionary columns, unlike previous work and uniform distance sampling.  This advantageous characteristic contributes to a notable decrease in the RMSE when localizing the grid point associated with a UE. This improvement is observed when comparing the performance of our proposed dictionary with the established benchmark methods.

\bibliographystyle{IEEEtran}
\bibliography{IEEEabrv,refs}

\end{document}